# Green tea induced gold nanostar synthesis mediated by Ag(I) ions

Qiang Chen[1,2], Toshiro Kaneko[1] & Rikizo Hatakeyama[1]

[1]Department of Electronic Engineering, Tohoku University, Sendai 980-8579, Japan,
[2]Present Address: Department of Bioengineering, The University of Tokyo, Tokyo, 113-8656, Japan
Email: chen@bionano.t.u-tokyo.ac.jp

## Abstract

We report a synthesis of tea components conjugated gold nanostars (AuNSs) with strong near infrared absorption by reducing an aqueous solution of chloroauric acid trihydrate via green tea in association with Ag(I) ions. Green tea acts as a reducing agent by providing electrons for the gold (III) reduction as well as a stabilizing agent by conjugating some of its components on the surfaces of AuNSs. Moreover, the Ag(I) ions play an important role in mediating the branched growth of the resultant AuNSs by inducing anisotropic growth on the surfaces of initially formed spherical gold nanoparticles.

# 1. Introduction

Gold nanoparticles (AuNPs) are attractive materials for both fundamental science and technological applications, such as surface-enhanced Raman spectroscopy (SERS), [1] molecular imaging, [2] sensors,[3] and photothermal therapy,[4-6] owing to their wide range of tunable surface plasmon resonance (SPR). The SPR of AuNPs is closely related to the size, shape, and assembly of AuNPs, [7] and therefore the control and mediation of these properties are major of interest on tuning the SPR in the AuNP research. Up to now, many types of AuNPs have been synthesized to tune the SPR of AuNPs, including sphere, polyhedra (such as cube and truncated cube), and nonpolyhedra (such as nanorod, nanoplate, and triangular plate).[8] Recently, researchers have succeeded in synthesizing gold nanostars (AuNSs) using both seeded [9-14] and seedless [15-18] growth methods. The strong near-infrared (NIR) absorptions of the AuNSs indicates potential applications of AuNSs in SERS,[19] catalysis,[19] and especially in photothermal therapy.[15, 20] However, the seeded growth method requires two separated steps that are similar to those commonly used in gold nanorod synthesis: a seeding process followed by a growth process. Each process requires elaborate reaction control,[9-14] making the seeded growth of AuNSs complicated and costly. Therefore, one-step seedless growth of AuNSs is preferred. Nonetheless, the reducing species used in both foregoing seeded and seedless methods are usually harmful to be used in vivo. In order to remove these disadvantages, a simple, low-cost, green synthesis of AuNSs is necessary.

Here we develop a simple, one-step green synthesis of AuNSs with strong broad NIR SPR absorptions by reducing an aqueous solution of chloroauric acid trihydrate via green tea in association with Ag(I) ions. The utilization of green tea not only decreases the cost, but also makes the AuNS synthesis as a green process due to its biocompatibility and non-toxicity.

## 2. Experimental

Our strategy is described as follows. X μl (X=0, 50, 70, 150, and 200) of silver nitrate (AgNO$_3$, $4\times10^{-3}$ M) was added to 1 ml of cetyltrimethylammonium bromide (CTAB, $1\times10^{-2}$ μM), and then 1 ml of chloroauric acid trihydrate (HAuCl$_4$·3H$_2$O, $1\times10^{-3}$ M) was added to the mixed solution. The solution quickly turned from colorless to brown yellowish. This brown yellowish solution immediately turned transparent after 1 ml of green tea was added. Moreover, the transparent solutions containing AgNO$_3$ gradually turned dark brown within 10 min, while solutions without AgNO$_3$ remained transparent for up to 20 min, and then gradually turned pink after 20 min. These results suggest that the presence of AgNO$_3$ accelerated the reaction process. However, we found that the growth rate of AuNPs was damped by the CTAB, because the formation of AuNPs occurred within several seconds in cases similar to our above experiment, but without CTAB. The whole synthesis process was performed at 25 ºC using de-ionized water as solvent. The mixed solutions were stood for 12 h after the green tea addition, and then we centrifuged the solutions and redispersed the precipitated products into de-ionized water for further characterization.

Commercially available bottled green tea (Oi Ocha Koiaji, Ito En, Ltd.) was used as purchased. Silver nitrate (AgNO$_3$, 99.5% in purity) and cetyltrimethylammonium bromide (CTAB, 98% in purity) were purchased from Wako Pure Chemical Industries, Ltd., while the chloroauric acid trihydrate (HAuCl$_4$·3H$_2$O, 99.9% in purity) was purchased from Sigma-Aldrich.

Samples for a TEM (Hitachi HF-2000) and EDX measurements were prepared by spreading a drop of AuNS solution onto a holey carbon film on a fine copper mesh and letting it dry. UV–vis absorption spectra were measured by a UV–vis spectroscopy (JASCO V-7200) using a standard quartz cell.

## 3. Results and discussion

Figure 1 shows transmission electron microscopy (TEM) images of the synthesized AuNPs as a function of the amount of $AgNO_3$ used. Evidently, an absence or excess of Ag(I) (X=0, 150, and 200 μl) inhibits the AuNS formation (Figs. 1a, 1e, and 1f), and as a result, spherical AuNPs are achieved. On the other hand, a few AuNSs were formed (see the scarce nanobranches in Fig. 1b) at 50 μl of $AgNO_3$, and AuNSs flourish at 70 μl of $AgNO_3$ (Fig. 1c). Finally, AuNS formation decreases as the amount of $AgNO_3$ increases to 100 μl (Fig. 1d).

Figure 2 presents the energy-dispersive X-ray (EDX) spectra at the core and the branch of the AuNSs corresponding to the sample in Fig. 1(c). It indicates that the core is mainly consisted with Ag or Ag coated Au and the branch is made from Au.

Figure 3 presents absorbance spectra of synthesized AuNPs. When no $AgNO_3$ is added, there is only a sharp SPR band at 550 nm, indicating the exclusive presence of spherical AuNPs, as shown in Fig. 1a. With 50 μl of $AgNO_3$, there is a weak broad NIR SPR band as well as a very weak SPR band at 545 nm (Fig. S1), indicating the presence of scarce nanobranches and large-sized AuNP aggregates, as shown in Fig. 1b. For AuNSs synthesized using 70 μl and 100 μl of $AgNO_3$, obvious SPR bands are observed at 545 nm and 810 nm, indicating the existence of AuNSs (Figs. 1c and 1d). A further increase in $AgNO_3$ amount results in a decrease of the absorbance and the disappearance of the SPR band at 810 nm, indicating the disappearance of AuNSs from the solutions as demonstrated in Figs. 1e and 1f.

Here, we attempt to elucidate the potential mechanism of AuNS formation by analyzing the reaction process. As described above, the immediate color change of solutions (CTAB/AgNO$_3$/HAuCl$_4$) from brown yellowish to transparent after adding green tea implies that the Au(III) is reduced to Au(0) as shown in reaction (1),

$$AuCl_4^-(aq) \xrightarrow{green\ tea} Au(s) + 4Cl^-(aq). \tag{1}$$

The Au(III) reduction is attributed to the electrons provided by catechins, [21, 22] and vitamins C [23] and E [24] which are constituents of green tea [25] (see Supplementary Information for details). At the very beginning, the Au(0) is in a monomer form, but it begins to nucleate and grow when the Au(0) concentration approaches a critical value [26] with continuous reaction (1), which results in the formation of initial spherical AuNPs. Subsequently, the initial spherical AuNPs act as basic cores for the growth of AuNSs. The presence of Ag(I) can result in the deposition of AgCl [27] and/or Ag(0) on the surfaces of the initial spherical AuNP cores during the synthesis, as shown in reactions (2),

$$\begin{aligned}Ag^+(aq) + Cl^-(aq) &\longrightarrow AgCl(s) \\ Ag^+(aq) &\xrightarrow{green\ tea} Ag(s)\end{aligned}. \tag{2}$$

Besides the reduction by green tea shown in reactions (2), Ag(I) can also be reduced to Ag(0) by underpotential deposition,[28] in which Ag(I) is reduced at a potential much lower than that required for bulk deposition. Because the first reaction of reactions (2) produces insoluble AgCl by consuming the Cl$^-$ product of reaction (1), the presence of AgNO$_3$ can inhibit the back-reaction of reaction (1).

This accelerates reaction (1), and thereby accelerates the AuNP formation, as observed in our experiment. The Au growth rate at sites with deposited AgCl and/or Ag(0) is decreased due to the protection of deposited hetero-materials, inducing anisotropic Au growth on the surfaces of the initial spherical AuNPs. A CTAB bilayer assembly might be formed on the AuNPs' surface via electrostatic interactions between the cationic CTAB head groups and anionic sites (absorbed $Cl^-$ and $Br^-$ or their related species) on the gold surface,[29] and the CTAB bilayer assembly on Au surface is more energetically favored on the nanorod than the nanosphere.[30] Therefore, the growth of gold nanobranches on the surfaces of the initial spherical AuNPs also benefits from the formation of the CTAB bilayer on the sides of the nanobranches, in a similar manner as in the Au nanorod synthesis.[29] Obviously, our results are consistent with the above proposed mechanism. In the absence or excess of Ag(I), the surfaces of the initial spherical AuNPs are Au atoms only or completely coated by AgCl and/or Ag(0), resulting in isotropic growth and the formation of spherical AuNPs. However, an intermediate amount of Ag(I) leads to partially inactivated AuNP surfaces which develop into AuNSs due to anisotropic growth on the surface of the initial spherical AuNPs.

In a synthesis of petal and spike-like AuNPs in the presence of poly(vinylpyrrolidone), [31] the petal and spike-like AuNPs were confirmed to mainly consist of Ag. Based on our results, however, we consider that our AuNSs consist of a core and branches as shown in the above analysis. The core is Ag-coated Au and the branches are mainly Au because the Ag content is small compared with Au content.

It is worth pointing out that green tea is not only a green reducing agent by providing electrons for Au(III) reduction, but also a stabilizing agent by conjugating some of its components on the surfaces of the synthesized AuNSs,[21-24] resulting in tea components conjugated AuNSs which are stable and compatible in vivo. The stability and compatibility in vivo renders the tea components conjugated AuNSs great advantages compared with other AuNSs usually with harmful conjugating materials, if used in practical disease therapy.

## 4. Conclusions

We present a simple, one-step, green synthesis of tea components conjugated AuNSs by reducing Au(III) via green tea in association with Ag(I) ions. Green tea acts as both reducing and stabilizing agents. Ag(I) is used to form inactive sites on the surfaces of initially formed spherical AuNPs to induce anisotropic Au growth on these surfaces. Water-soluble, tea components conjugated AuNSs can be formed in solutions with intermediate amounts of Ag(I), but not with excessive or insufficient Ag(I). Moreover, the synthesized AuNSs show strong broad NIR absorptions, which enables significantly possible applications in cancer cell imaging and destruction, and in photothermal therapy.[4, 5, 32, 33]

## Acknowledgments

We thank the financial supports from Japan Society for the Promotion of Science of Postdoctoral Fellowship for Foreign Researchers and Tohoku University 21st Century Center of excellence Program.

**Figure captions**

**Figure 1.** TEM images of synthesized AuNPs as a function of added AgNO$_3$ amount. (a)-(f) correspond to 0, 50, 70, 100, 150, and 200 µl of AgNO$_3$, respectively. Scale bar is 50 nm.

**Figure 2.** Energy-dispersive X-ray spectra at the core and the branch of the gold nanostars. Element is indicated in the figure. Sample was made by adding 70 µl AgNO$_3$ (4×10$^{-3}$ M) into 1 ml of CTAB (1×10$^{-2}$ µM), and then 1 ml of HAuCl$_4$·3H$_2$O (1×10$^{-3}$ M) was added into the mixture.

**Figure 3.** Absorbances of synthesized AuNPs as a function of added AgNO$_3$ amount.

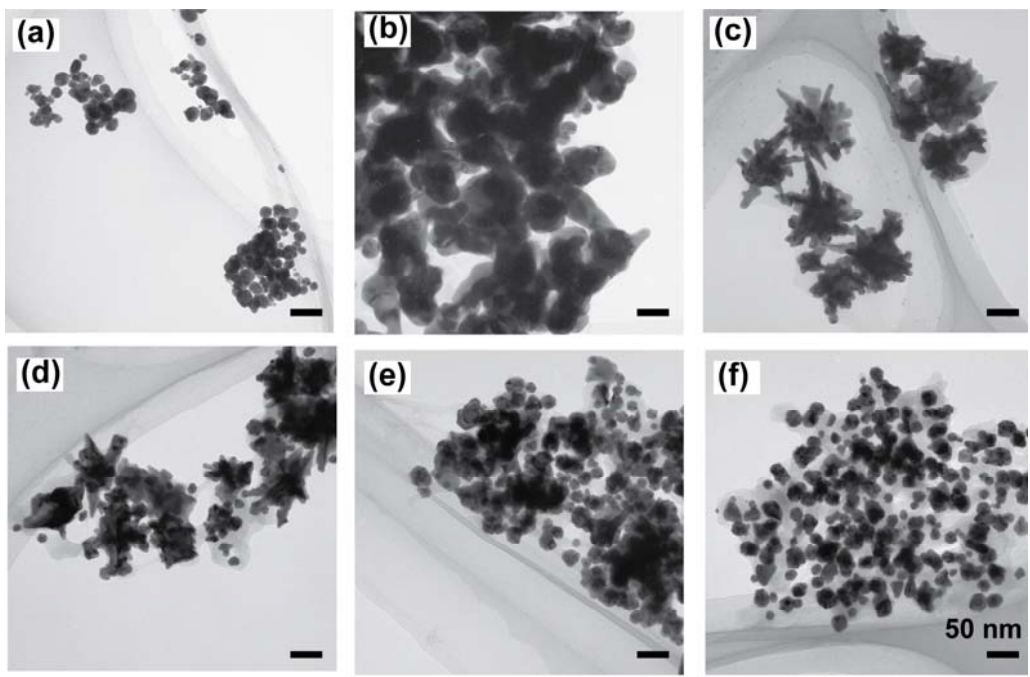

Fig. 1

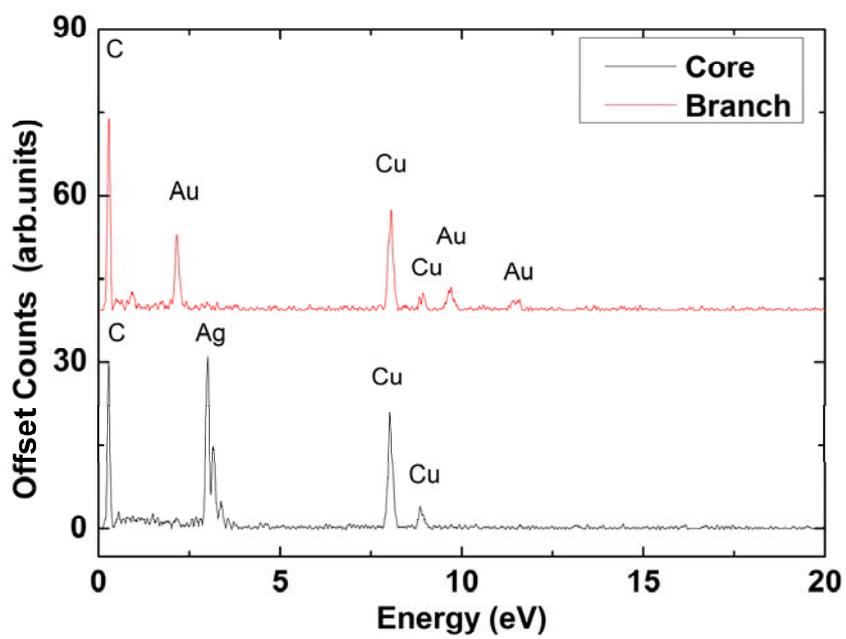

Fig. 2

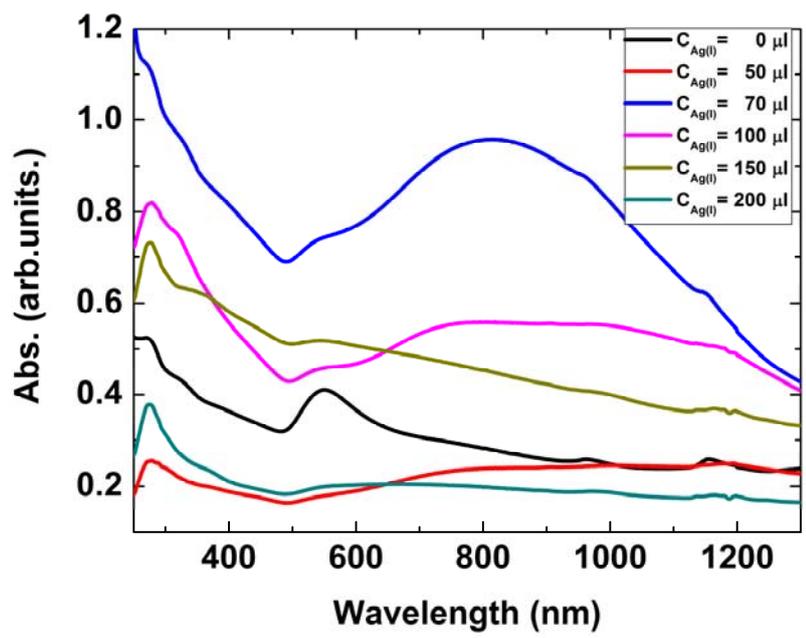

Fig. 3

# Electronic Supplementary Information for
# Green tea induced gold nanostar synthesis mediated by Ag(I) ions

Qiang Chen[1,2], Toshiro Kaneko[1] & Rikizo Hatakeyama[1]

[1]Department of Electronic Engineering, Tohoku University, Sendai 980-8579, Japan,
[2]Present Address: Department of Bioengineering, The University of Tokyo, Tokyo, 113-8656, Japan

Correspondence and requests for materials should be addressed to Q.C. (chen@bionano.t.u-tokyo.ac.jp)

## 1. Chemicals used in the experiments

Commercially available green tea (Oi Ocha Koiaji, Ito En, Ltd.) was used as purchased for reducing species, consisting catechins (0.8 g/l) and sodium (0.92 g/l) according to Ref. [34]. Silver nitrate ($AgNO_3$, 99.5% in purity) and cetyltrimethylammonium bromide (*CTAB*, 98% in purity) were purchased from Wako Pure Chemical Industries, Ltd., while the chloroauric acid trihydrate ($HAuCl_4·3H_2O$, 99.9% in purity) was purchased from Sigma-Aldrich.

## 2. Magnified absorbance of gold nanoparticles (*AuNPs*) synthesized with 50 μl of *$AgNO_3$*

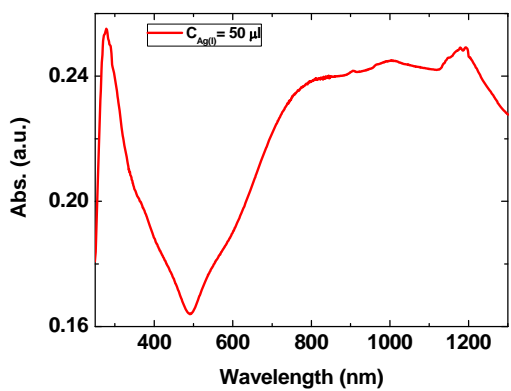

**Figure S1.** Absorbance of *AuNP*s synthesized with 50 μl of *$AgNO_3$*.

## 3. Mechanism of *Au*(III) reduction by green tea

Green tea contains many components (catechins, Vitamins C and E etc.) [25], and some of them are demonstrated to have reducing abilities. For example, a simple form of flavonoid known as catechins can reduce *Cu* [35], *Fe* [35], and *Au* [21, 22] ions, and vitamins *C* [23] and *E* [24] can reduce *Ag(I)* to *Ag(0)*. In green tea, since the process for making crude tea involves halting the action of oxidizing enzymes (crude tea refers to tea leaves that have had the action of oxidizing enzymes halted, fermentation prevented, had their moisture content reduced somewhat and processed into a state able to withstand storage.), most of the catechins remain unoxidized, unlike in oolong and black teas in which they are mostly oxidized [36]. These unoxidized catechins retain their reducing capability. The formula of a catechin is shown in Fig. S2. Catechins have two benzene rings (called the A- and B-rings) and a dihydropyran heterocycle (the C-ring) with a hydroxyl group on carbon 3 [37]. There are two chiral centers on the molecule at carbons 2 and 3, and there are therefore four diastereoisomers. During reaction, electron donation initially occurs very easily and reversibly at the two hydroxyl groups in the B-ring, with a very low positive potential. Subsequently, electron donations occur at hydroxyl groups in the A-ring by an irreversible process. In our case, the reduction of *Au(III)* is attributed to electrons donated by the catechins, and/or Vitamins C and E in green tea.

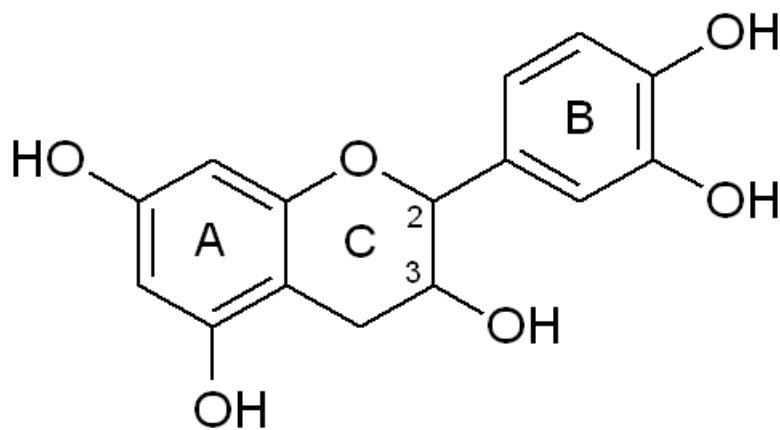

**Figure S2.** Catechin.